\documentclass[journal]{IEEEtran}

\usepackage{hyperref}
\hypersetup{hidelinks}
\usepackage{xcolor,soul,framed} 

\colorlet{shadecolor}{yellow}
\usepackage[pdftex]{graphicx}
\DeclareGraphicsExtensions{.pdf,.jpeg,.png, .jpg }

\usepackage{amsfonts,amssymb}
\usepackage{mathrsfs}
\usepackage{amsthm}
\usepackage[cmex10]{amsmath}
\usepackage{mathtools}
\usepackage{bbm}
\DeclareMathAlphabet{\mathbbb}{U}{bbold}{m}{n}
\usepackage{array}
\usepackage{flushend}
\usepackage{physics}
\usepackage{cite}
\usepackage{comment}

\theoremstyle{remark}
\newtheorem{theorem}{Theorem}

\newtheorem{proposition}{Proposition}

\newtheorem{remark}{Remark}

\newcommand{\ie}{i.e.}

\newcommand{\pha}[1]    {\underline{#1}}

\begin{document}
\bstctlcite{IEEEexample:BSTcontrol}
    \title{ Voltage Synchronization and Proportional Current Sharing of Grid-Forming Inverters}
    \author{Qianxi~Tang,~\IEEEmembership{Student Member,~IEEE,}
          and~Li~Peng,~\IEEEmembership{Senior Member,~IEEE}

  \thanks{This work has been submitted to the IEEE for possible publication.
  	
  	The authors are with School of Electrical and Electronic Engineering, HUST, Wuhan 430074, China. Email: qianxi@hust.edu.cn, pe105@mail.hust.edu.cn}}

\maketitle


\begin{abstract}
Most previously proposed controllers are analyzed in the small-signal/quasi-steady regime rather than large-signal or transient stability for grid-forming inverters (GFMI). Additionally, methods that presume system-wide data—global measurements and complete grid-model knowledge—are challenging to realize in practice and unsuitable for large-scale operation. Moreover, proportional current sharing is rarely embedded into them. The whole system is a high-order, nonlinear differential system, making analysis intractable without principled simplifications. 
Hence, contraction stability analysis in GFMI is proposed to guarantee the large-signal stability. Furthermore, a contraction-based controller is proposed to synchronize GFMI. Additionally, this paper proposes integrating an auxiliary virtual-impedance layer into the contraction-based controller to achieve proportional current sharing, while the GFMI retains global stability and voltage synchronization. A dispatchable virtual oscillator control (dVOC), also known as the Andronov--Hopf oscillator (AHO) is used to validate the proposed contraction stability analysis and contraction-based controller with virtual-impedance.
It is proved that the complex multi-converter system can achieve output-feedback contraction under \textit{large-signal} operation. Therefore, without requiring system-wide data, the proposed method offers \textit{voltage synchronization}, \textit{decentralized stability conditions} for the \textit{transient stability} of AHO and  \textit{proportional current sharing}, beyond prior small-signal, quasi-steady analysis. 
\end{abstract}


\section{Introduction}

\IEEEPARstart{R}{enewable} energy sources are integrated to power system. Unlike large-capacity synchronous generators, renewable power generation features a substantial number of small-capacity generating units without mechanical inertial \cite{liu2022stability}.  Conventional stability analysis faces aggregated equivalent modeling and analysis \cite{li2017practical}, which is rarely applicable. Therefore, decentralized controls without communication and their stability analysis are principally preferred. The time-domain virtual oscillator control (VOC), culminating in dispatchable VOC schemes with Andronov--Hopf oscillators (AHO/dVOC). AHOs generate harmonic-free limit cycles with nonlinear droop-like behavior and autonomous synchronization using only local measurements, avoiding PLLs and explicit power calculations while retaining fast dynamics. Surveys and models of AHO/dVOC document these properties and their realization in practical three-phase inverters. At the same time, Hopf-versus Van der Pol comparisons report faster recovery, reduced harmonics, and improved robustness for Hopf oscillators under identical conditions, with global asymptotic synchronization shown via Lyapunov arguments for islanded parallel inverters \cite{Hopf2021}. In contrast, droop/VSG schemes can suffer from oscillatory active-power transients and inaccurate reactive-power sharing unless carefully augmented \cite{Quantitative2024,Synchronization2022}.

While AHO/dVOC offers attractive performance and a rich small-signal theory, uniform guarantees that hold globally in state space and nonlinearly under load disturbances, model mismatch, and topology changes remain limited. Classical Lyapunov or small-signal analyses provide local stability and modal insight but typically hinge on equilibrium linearization, restrictive coupling conditions, or specific operating points \cite{osti_2418742}. Additionally, in multi–converter grids the states are tightly coupled through the  network dynamics : a change at one unit instantaneously
perturbs the PCC voltage seen by all others, yielding a high–order, nonlinear
coupled system that is hard to analyze at large signal, which makes globally stable proof  more difficult \cite{Transmission2019}. While some studies assess multi-converter stability using system-wide data—complete topology, load conditions, and line impedances—such information is rarely available or maintainable in practice, making these methods impractical at scale \cite{Kron2013}.
 Moreover, when virtual impedance is introduced to limit transient currents, there is a need for guarantees that this augmentation does not compromise synchronization or stability \cite{Limiting2022}. In multi-converter systems it remains challenging to guarantee that droop-like and AHO virtual-oscillator controls ensure large-signal and transient stability, voltage synchronization, and accurate current sharing; most prior results largely address  small-signal stability or quasi-steady regimes \cite{Hopf2021,Passivity2024}.

Contraction theory provides large-signal, time-varying stability certificates by
showing that distances between trajectories decay exponentially, applicable to both nonlinear and linearized systems \cite{wang2005partialcontraction}. Contraction analysis adopts a different viewpoint on stability: instead of asking whether trajectories approach a particular equilibrium or nominal motion, it asks whether distances between trajectories shrink over time. If nearby solutions converge to one another (in a suitable metric), the system progressively “forgets’’
its initial conditions and temporary disturbances; the eventual behavior is then
independent of how the system was started. This leads to a differential
stability test—based on local Jacobian/metric properties—rather than searching for
a global motion integral as in classical Lyapunov arguments or relying on a global
state transformation as in feedback linearization \cite{Contraction2006,Contraction2000}. The result is a powerful and
often simpler framework for nonlinear, time-varying systems. Contraction
yields global, incremental stability via partial contraction
for synchronization, and admits decentralized checks through local Jacobian/metric
inequalities—making it a powerful analysis tool for scalable integration of
grid-forming inverters. However, to the best of our knowledge, a  decentralized contraction-theoretic framework for grid-forming inverters has not been reported; this is the theoretical contribution of our study and introduces contraction as a scalable analysis tool for large-scale integration.

This paper proposes contraction stability analysis in GFMI and a contraction-based controller with virtual-impedance  to analyze and regulate the dynamics of AHO. The contraction stability analysis certifies \emph{global asymptotic stability with exponential convergence} and \emph{proportional current sharing} \emph{without} quasi-steady assumptions or system-wide data. The entire guarantee reduces to a simple, fully decentralized \emph{algebraic} inequality.

The remainder of this paper is organized as follows. In Section II, mathematical notations used in this paper and the contraction concepts will be illustrated with some related theorems. In Section III, the proposed  contraction-based control with virtual-impedance is constructed based on the previous concepts and theorems.  The voltage synchronization and proportional current sharing in large-signal operation will be proved. Also, a simple, fully decentralized algebraic inequality of contraction stability condition is derived.  Section IV Validation of contraction stability analysis and contraction-based controller with virtual-impedance for GFMI  based on simulation of a real wind power plant is presented. Finally, Section V concludes this paper.

\begin{figure}
	\begin{center}
		\includegraphics[width = 1\linewidth]{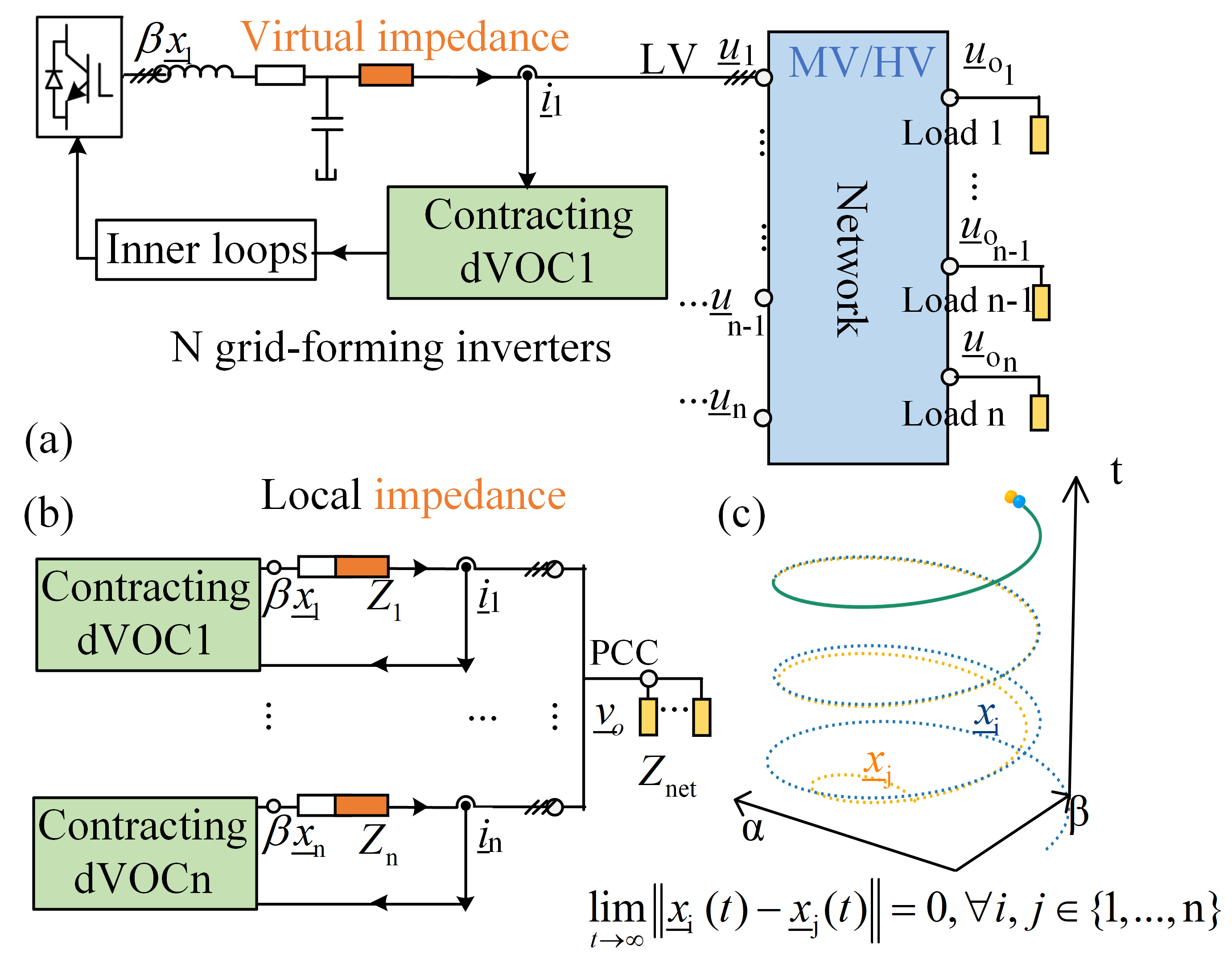}
		\caption{ The studied GFMI diagram with state synchronizing. (a) Multi-converter grid-connected systems with proposed contracting dVOC of grid-forming. (b) System model with feedback connection when load/local-impedance dominate the system’s voltage drop. (c) Voltage trajectories of n contracting inverters in \textit{$\alpha\beta$ reference frame} over time.} 
		\label{fig:block-diagram}
	\end{center}
\end{figure}

\section{A Novel Contraction Stability Analysis for Grid-forming Inverters}
\label{sec:passivity analysis}


In this section, the contraction concepts will be illustrated with some corresponding theorems which will be adopted to construct contraction-based control with virtual impedance  in Section III. It can be shown that even with modeling error or deterministic disturbance, the contraction guarantees all the trajectories of grid-forming inverters converge to a particular solution exponentially with a bounded steady-state error .
\subsection{Notation}
\label{notation}
For a square matrix $A^{n\times n}$, we use the notation $A \succ 0$, $A \succeq 0$, $A \prec 0$, and $A \preceq 0$ for the positive definite, positive semi-definite, negative definite, negative semi-definite matrices, respectively.  Furthermore, we use $f_{{\mathbf{x}}} = \partial f/\partial {\mathbf{x}}$, $M_{x_i} = \partial M/\partial x_i$, and $M_{x_ix_j} = \partial^2 M/(\partial x_i\partial x_j)$, where $x_i$ and $x_j$ are the $i$th and $j$th elements of ${\mathbf{x}} \in \mathbb{R}^n$, for describing partial derivatives in a limited space. The other notations are given in Table~\ref{tab:notations_in_this_paper}. 

In our modeling and analysis for GFMI, a three-phase balanced condition is considered and the \textit{$\alpha\beta$ reference frame} is employed \cite{osti_2418742} without  assuming synchronized frequency. As shown in Fig.~\ref{fig:block-diagram}, the converter \textit{voltage} and the \textit{output current} are expressed as $\pha x_k \coloneqq x_{{\rm \alpha},k} +j x_{{\rm \beta},k}=(\Re\{\pha x_k\},\,\Im\{\pha x_k\})^{\!\top} $ and $\pha i_k \coloneqq i_{{\rm \alpha},k} +j i_{{\rm \beta},k}=(\Re\{\pha i_k\},\,\Im\{\pha i_k\})^{\!\top}$, respectively. This is the bridge between the complex variables and vector space variables. Underlines are used throughout to indicate complex variables.

\begin{table}
	\caption{Notations used in this paper. \label{tab:notations_in_this_paper}}
	\footnotesize
	\begin{center}
		\renewcommand{\arraystretch}{1.2}
		\begin{tabular}{ |c|m{6.3cm}| } 
			\hline
			$\|{\mathbf{x}}\|$ & Euclidean norm of $x \in \mathbb{R}^n$ \\ \hline
			$\delta {\mathbf{x}}$ & Differential displacement of $x \in \mathbb{R}^n$ \\ \hline
			$\|A\|$ & Induced $2$-norm of $A \in \mathbb{R}^{n\times m}$ \\ \hline
			$\lambda_{\min}(A)$  & Minimum eigenvalue of $A\in\mathbb{R}^{n\times n}$ \\ \hline
			$\lambda_{\max}(A)$  & Maximum eigenvalue of $A\in\mathbb{R}^{n\times n}$ \\ \hline $\mathrm{I}$ & Identity matrix of appropriate dimensions \\ \hline
			$\mathbb{R}_{>0}$ & Set of positive reals, \ie{}, $\{a\in\mathbb{R}|a\in(0,\infty)\}$ \\ \hline
			$\mathbb{R}_{\geq 0}$ & Set of non-negative reals, \ie{}, $\{a\in\mathbb{R}|a\in[0,\infty)\}$ \\ \hline
		\end{tabular}
	\end{center}
\end{table}
\subsection{Contraction Theorems }

Consider the smooth non-autonomous system
\begin{equation}
	\label{eq:xfx}
	\dot{\mathbf{x}}(t)=f\bigl(\mathbf{x}(t),t\bigr),
\end{equation}
where $t \in \mathbb{R}_{\geq 0}\; \text{is time}, \text{states}\; \mathbf{x}:\mathbb{R}_{\ge0}\!\to\mathbb{R}^n,\;
f:\mathbb{R}^n\times\mathbb{R}_{\ge0}\!\to\mathbb{R}^n$
whose smoothness ensures local existence and uniqueness 
of the solution to \eqref{eq:xfx} for a given $\mathbf{x}(0)=\mathbf{x}_0$ at least locally. Let $N$ copies evolve as
$
\dot{\mathbf{x}}_i(t)=f\bigl(\mathbf{x}_i(t),t\bigr),\; \mathbf{x}_i(t)\in\mathbb{R}^n,\; i=1,\dots,N.
$

\begin{theorem}[Contracting \cite{Contraction2021}]
	\label{Thm:contraction}
	If there exists a uniformly positive definite matrix ${M}({\mathbf{x}},t)={{\Theta}}({\mathbf{x}},t)^{\top}{{\Theta}}({\mathbf{x}},t) \succ 0,~\forall \mathbf{x},t$, where ${\Theta(\mathbf{x},t)}$ defines a smooth coordinate transformation of $\delta x$, \ie, $\delta{z}={\Theta}(\mathbf{x},t)\delta{\mathbf{x}}$, \st{} either of the following equivalent conditions holds for $\exists \alpha \in \mathbb{R}_{>0}$, $\forall \mathbf{x},t$:
	\begin{align}
		&\lambda_{\max}(F(\mathbf{x},t))=\lambda_{\max}\left(\left({\dot{\Theta}}+{{\Theta}}\frac{\partial
			{f}}{\partial
			{\mathbf{x}}}\right){{\Theta}}^{-1}\right) \leq - \alpha
		\label{eq_MdotContracting_z} \\
		&{\dot{M}}+M\frac{\partial
			{f}}{\partial {{\mathbf{x}}}}+\frac{\partial {f}}{\partial
			{\mathbf{x}}}^{\top}M \preceq -2\alpha M
		\label{eq_MdotContracting}
	\end{align}
	where the arguments $(\mathbf{x},t)$ of $M(\mathbf{x},t)$ and $\Theta(\mathbf{x},t)$ are omitted for notational simplicity, then all the solution trajectories of \eqref{eq:xfx} converge to a single trajectory exponentially fast regardless of their initial conditions (\ie{}, contracting), with an exponential convergence rate $\alpha$. The converse also holds.
\end{theorem}
\begin{theorem}[Robustness under perturbation \cite{Contraction2021}]
	\label{thm:incstab_robust}
	Consider $\dot {\mathbf{x}} = f( {\mathbf{x}},t)$ and suppose there exists a smooth Riemannian metric
	$M( {\mathbf{x}},t)=\Theta({\mathbf{x}},t)^{\!\top}\Theta({\mathbf{x}},t)\succ 0$ and $\alpha>0$ such that
	\begin{equation}
		\dot M + M\frac{\partial f}{\partial \mathbf{x}} + \frac{\partial f}{\partial {\mathbf{x}}}^{\!\top} M \;\preceq\; -\,2\alpha\,M
		\quad \text{for all }({\mathbf{x}},t).
		\label{eq:contr_cond}
	\end{equation}
	Let $\xi_0(t)$ and $\xi_1(t)$ be any two solutions of $	\dot{\mathbf{x}}(t)=f\bigl(\mathbf{x}(t),t\bigr)$ and let
	\[
	V_\ell(t) \;:=\; \inf_{\text{paths }{\mathbf{x}}(\mu,t)} \int_{0}^{1} \bigl\|\Theta({\mathbf{x}},t)\,\partial_\mu x\bigr\|\,d\mu
	\]
	be the Riemannian path length (geodesic distance) between $\xi_0(t)$ and $\xi_1(t)$.
	Then
	\begin{equation}
		V_\ell(t) \;\le\; e^{-\alpha t}\,V_\ell(0),
		\qquad
		\|\xi_1(t)-\xi_0(t)\| \;\le\; \frac{e^{-\alpha t}}{\sqrt{\underline m}}\,V_\ell(0),
		\label{eq:inc_exp_bound}
	\end{equation}
	whenever $M({\mathbf{x}},t)\succeq \underline m I$.
	Hence, any two trajectories converge exponentially (incremental stability).
	
	Now consider the perturbed system $\dot {\mathbf{x}}=f({\mathbf{x}},t)+d({\mathbf{x}},t)$ with $\|d(x,t)\|\le\bar d$ and
	assume $\underline m I \preceq M({\mathbf{x}},t)\preceq \overline m I$.
	Then the geodesic distance and the Euclidean separation satisfy
	\begin{align}
		V_\ell(t)
		&\le e^{-\alpha t}V_\ell(0)\\
	    &+ \frac{\sup_{{\mathbf{x}},t}\|\Theta({\mathbf{x}},t)\,d({\mathbf{x}},t)\|}{\alpha}\,\bigl(1-e^{-\alpha t}\bigr), \\
		\|\xi_1(t)-\xi_0(t)\|
		&\le \frac{V_\ell(0)}{\sqrt{\underline m}}\,e^{-\alpha t}
		+ \frac{\bar d}{\alpha}\sqrt{\frac{\overline m}{\underline m}}\,(1-e^{-\alpha t}).
		\label{eq:robust_ball}
	\end{align}
	
	Consequently, the separation decays exponentially and remains bounded by a disturbance-to-state “error ball” of radius
	$\displaystyle \frac{\bar d}{\alpha}\sqrt{\tfrac{\overline m}{\underline m}}$.
\end{theorem}

\begin{theorem}[Partial Contraction Based Synchronization \cite{Slotine2005}]
	\label{thm:sync}
	Consider $N$ vector systems with states $\mathbf{x}_i(t)\in\mathbb{R}^n$.
	If there exists a \emph{contracting} vector field
	$\mathbf{h}:\mathbb{R}^n\times\mathbb{R}_{\ge0}\!\to\mathbb{R}^n$ such that
	\[
	\dot{\mathbf{x}}_1 - \mathbf{h}(\mathbf{x}_1,t)
	= \cdots =
	\dot{\mathbf{x}}_N - \mathbf{h}(\mathbf{x}_N,t),
	\]
	then all trajectories converge exponentially regardless of the initial conditions. Namely,  $\forall\, i,j \in i=1,\dots,N$,
	\[
	\|\mathbf{x}_i(t)-\mathbf{x}_j(t)\|
	\le \exp\!\left(\int_{0}^{t}\lambda_{\max}\big(\mathbf{x},\tau\big)\,d\tau\right)\,\|\mathbf{x}_i(0)-\mathbf{x}_j(0)\|,
	\]
	where the convergence rate is $\lambda_{\max}(\mathbf{x},t)$ which is the largest
	eigenvalue of the symmetric part of the Jacobian
	$J=\frac{\partial \mathbf{h}}{\partial \mathbf{x}}$, i.e.,
	$\tfrac{1}{2}\!\left(J+J^{\top}\right)$. Hence, if
	$\lambda_{\max}(\mathbf{x},t)$ is uniformly strictly negative, any $\|\mathbf{x}_i(t)-\mathbf{x}_j(t)\|$ converges exponentially to zero.
\end{theorem}
\paragraph*{Scope (large-signal, high order)}
Although contraction analysis contain many other theorems, these theorems are used to certify \emph{large-signal} (nonlinear, time-varying) stability for GFMI—no linearization or quasi-steady approximation is required. This paper is trying to construct a contraction-theoretic stability analysis for the synchronization of GFMI.

\paragraph*{Application requirement (how we use it)}
To invoke theorems of Section II in multi-inverter grids, each inverter’s closed-loop dynamics
must be cast into the \emph{partial-contraction} form
$\dot{\mathbf{x}}_i-\mathbf{h}(\mathbf{x}_i,t)=\mathbf{r}(t)$ with the same
right-hand side $\mathbf{r}(t)$ for all $i$. Achieving this form via
contraction-based control and enforcing a uniform
inequality that makes $\mathbf{h}$ contracting constitute the key, \emph{novel}
modeling step. This step turns a complex networked nonlinear problem into a decentralized,
algebraic certificate for large-signal synchronization and current sharing.

\section{ System Dynamics With the Proposed Contraction-Based Controller}

Grid-forming converters measure their output currents and establish the terminal voltages, as depicted in Fig.~\ref{fig:block-diagram}(a).   Each inverter is on the LV side and is stepped up by a transformer to the MV/HV network where loads are connected. Usually, loads consume the currents that produce the voltage drop across themselves. Compared to loads, the MV/HV network’s own impedance is small, so its drop is negligible. Furthermore, at each inverter branch, the local series impedance (virtual and local line) can be designed higher than the network impedance in pu unit. So, it is said that local impedance and load dominate the whole system voltage drop (domination condition) and MV/HV network has negligible voltage drop as shown in Fig.~\ref{fig:block-diagram}(b), which will be backed up in the 
Validation, Section~IV.


\textit{1) Synchronized Network Representation:}

\paragraph{Current sharing for $n$ parallel inverters}
Let inverter $i\in\{1,\dots,n\}$ be modeled by an internal voltage $\beta \pha x_i = E_i e^{j\theta_i}$, behind the series local impedance $Z_i$; let the downstream network seen from the PCC be $Z_{\mathrm{net}}$ as shown in Fig.~\ref{fig:block-diagram}(b). Define admittances
\begin{equation*}
	\begin{aligned}
	&Y_i:=1/Z_i, Y_{\mathrm{net}}:=1/Z_{\mathrm{net}}\\
	&Y_\Sigma \;:=\; \sum_{m=1}^{n} Y_m \;+\; Y_{\mathrm{net}}.
	\end{aligned}
\end{equation*}

The PCC gives the bus voltage $\pha v_o(t) =V $.
KCL at the PCC gives the bus voltage and inverter currents:
\begin{align}
	V &= \frac{\displaystyle \sum_{m=1}^{n} \frac{E_m e^{j\theta_m}}{Z_m}}{\displaystyle \sum_{m=1}^{n}\frac{1}{Z_m} + \frac{1}{Z_{\mathrm{net}}}}
	\,=\, \frac{\displaystyle \sum_{m=1}^{n} Y_m E_m e^{j\theta_m}}{Y_\Sigma},\\[4pt]
	I_i &= \frac{E_i e^{j\theta_i}-V}{Z_i}
	\,=\, Y_i E_i e^{j\theta_i} \;-\; \frac{Y_i}{Y_\Sigma}\sum_{m=1}^{n} Y_m E_m e^{j\theta_m}.
\end{align}

Hence the exact sharing ratio between units $i$ and $j$ is
\[
\frac{I_i}{I_j} \;=\; 
\frac{Y_i\!\left(E_i e^{j\theta_i}-V\right)}{Y_j\!\left(E_j e^{j\theta_j}-V\right)},
\qquad
V=\frac{\sum_{m} Y_m E_m e^{j\theta_m}}{Y_\Sigma}.
\]

 Hence, for \emph{Identical $\beta \pha x_i$ (synchronized case):} if $E_i e^{j\theta_i}=E e^{j\theta}$ for all $i$, then
\begin{equation}
\label{eq:finalsteady}
V = E e^{j\theta}\,\frac{\sum_{m}Y_m}{Y_\Sigma},\qquad
I_i = E e^{j\theta}\,\frac{Y_i\,Y_{\mathrm{net}}}{Y_\Sigma},
\end{equation}
so the current-sharing ratio is independent of the downstream network:
$
\;I_1:I_2:\cdots:I_n \;=\; \frac{1}{Z_1}:\frac{1}{Z_2}:\cdots:\frac{1}{Z_n}.\;
$

\paragraph{The Solution of N Converged Inverters system}
When all the $\beta \pha x_i$ converged, using the equivalent circuit transformation, the n inverters can be seen as all connected at a common point. So, the $\pha v_o$ can be obtained. According to \eqref{eq:finalsteady}, $\pha v_o(t) = \beta \pha x_i\,\frac{\sum_{m}Y_m}{Y_\Sigma}$, the admittance can be regarded as a filter. 
\begin{equation}
\pha v_o = \beta \pha x_i\,\frac{\sum_{m}Y_m}{Y_\Sigma}.
\end{equation}
When $\sum_{m}Y_m >> Y_{\mathrm{net}}$, the $ \pha v_o(t) \;\approx\; \beta\,\pha x_i$.

\textit{2) Grid-Forming Dynamics With Contraction-Based Controller With Virtual-Impedance:}  

After transformed into \textit{$\alpha\beta$ reference frame}, the dVOC grid-forming voltage dynamics are given as \cite{osti_2418742}
\begin{equation}
	\label{eq:aho_ss}
	\begin{aligned}
		\dot x_{{\rm \alpha},k}&=\chi\,x_{{\rm \alpha},k}-\omega_0 x_{{\rm \beta},k},\\
		\dot x_{{\rm \beta},k}&=\chi\,x_{{\rm \beta},k}+\omega_0 x_{{\rm \alpha},k},
	\end{aligned}
\end{equation}
 where $\chi=\xi\!\left(2X_{\rm nom}^2-x_{{\rm \alpha},k}^2-x_{{\rm \beta},k}^2\right).$ $\chi$ is a nonlinear, state-dependent scalar that varies with the Euclidean norm  $\|\pha x_k\|$. It yields oscillations with root mean square (rms) amplitude
 $X_{\mathrm{nom}}$ regardless of initial conditions. In addition,
 $\xi$ is a constant that dictates the convergence speed to a limit cycle. Fig.~\ref{fig:block-diagram}(c) sketches trajectories yielded by~\eqref{eq:aho_ss}: the state $\pha x_k$ trajectory always spiral asymptotically toward the
 stable circular limit cycle with a fixed radius $\sqrt{2}\,X_{\mathrm{nom}}$
 and constant rotation frequency $\omega_0$. However, to realize $ \forall\, i,j \in \{1,\dots,n\}:\;\lim_{t\to\infty}\|\pha x_i(t)-\pha x_j(t)\|=0$, contracting converters is needed.
 
The contraction-based controller with virtual-impedance is designed to do it. For each inverter, if the local control is  $\kappa \,f(\pha i_k(t)),\,k \in \{1,\dots,n\}, \kappa$ is a constant gain, this gets
\begin{equation}
	\label{eq:aho_ssc}
		\begin{aligned}
			\dot x_{{\rm \alpha},k}&=\chi\,x_{{\rm \alpha},k}-\omega_0 x_{{\rm \beta},k}+\kappa \,f(i_{\alpha,k}(t)),\\
			\dot x_{{\rm \beta},k}&=\chi\,x_{{\rm \beta},k}+\omega_0 x_{{\rm \alpha},k}+\kappa \,f(i_{\beta,k}(t)).
	\end{aligned}
\end{equation}

The voltage drop across the network is negligible. Based on Fig.~\ref{fig:block-diagram}(b), for each converter branch after adding the virtual impedance  $ r_v + j X_v$, this has
\begin{equation*}
	\label{eq:model-alpha}
	L_{\rm f,k}\,\dot {\pha i}_k(t)+  r_{\rm f,k}\, \pha i_k(t)+\big(r_v + j X_v\big)\,\pha i_k(t)= \beta\, \pha x_k-  \pha v_o(t) ,
\end{equation*}
 where $\beta$ is a constant real number. $ L_{\rm f,k}$  and $r_{\rm f,k}$ are the known local line parameters. 
 
So, define
 \begin{align*}
  \,f(\pha i_k(t)) \;:=\;- \!\left(L_{\rm f,k}\,\dot {\pha i}_k(t)+  r_{\rm f,k}\, \pha i_k(t)+\big(r_v + j X_v\big)\,\pha i_k(t)\right).
 \end{align*}

As a result, each grid-forming dynamics becomes
\begin{equation}
\label{eq:eachADF}
\begin{aligned}
	\dot x_{{\rm \alpha},k} &= \chi\,x_{{\rm \alpha},k} - \omega_0 x_{{\rm \beta},k} - \kappa\!\left(\beta\,x_{{\rm \alpha},k} -v_{o,\alpha}(t)\right),\\[2pt]
	\dot x_{{\rm \beta},k} &= \chi\,x_{{\rm \beta},k} + \omega_0 x_{{\rm \alpha},k} - \kappa\!\left(\beta\,x_{{\rm \beta},k} - v_{o,\beta}(t)\right).
\end{aligned}
\end{equation}

Now, with this contraction controller, it is ready to use the contraction analysis to guarantee the synchronization and current sharing.

\begin{proposition}
	\label{prop:network}
	For \eqref{eq:eachADF}, if$\;\kappa\beta-2\xi X_{\rm nom}^{2} \ge \displaystyle c>0\;$, 
the n inverters will achieve  $ \forall\, i,j \in \{1,\dots,n\}:\;\lim_{t\to\infty}\|\pha x_i(t)-\pha x_j(t)\|=0$.
\end{proposition}

\begin{proof}
Let
\begin{equation*}
\mathbf{x}_k^\top:=(x_{\alpha,k},x_{\beta,k}),\quad
J:=\begin{bmatrix}0&-1\\[2pt]1&0\end{bmatrix},\quad I:=\begin{bmatrix}1&0\\0&1\end{bmatrix}.
\end{equation*}

 Then, 
\begin{equation*}
\dot{\mathbf{x}}_k
= \big(\chi I+\omega_0 J-\kappa\beta I\big)\,\mathbf{x}_k
+ \kappa\,\pha v_o(t).
\end{equation*}
 Define the contracting local map
\begin{equation*}
\;\mathbf{h}(\mathbf{x},t):=\big(\chi I+\omega_0 J-\kappa\beta I\big)\,\mathbf{x},\;
\end{equation*}
so that
\begin{equation*}
\;\dot{\mathbf{x}}_k-\mathbf{h}(\mathbf{x}_k,t)=\kappa\,\pha v_o(t),\;\forall k \in \{1,\dots,n\}\;
\end{equation*}
 Its Jacobian (${\Theta}(x,t)= I $) is
\begin{equation*}
\frac{\partial \mathbf{h}}{\partial \mathbf{x}}
= \big(\chi-\kappa\beta\big)I+\omega_0 J-2\xi\,\mathbf{x}\mathbf{x}^{\!\top}.
\end{equation*}
The symmetric part (Euclidean metric) is
\begin{equation*}
\tfrac{1}{2}\!\left(\frac{\partial \mathbf{h}}{\partial \mathbf{x}}
+\frac{\partial \mathbf{h}}{\partial \mathbf{x}}^{\!\top}\right)
=\big(\chi-\kappa\beta\big)I-2\xi\,\mathbf{x}\mathbf{x}^{\!\top}
\;\preceq\; \big(2\xi X_{\rm nom}^2-\kappa\beta\big) I .
\end{equation*}

Hence, the $\lambda_{\max}(\mathbf{x},t)\le2\xi X_{\rm nom}^2-\kappa\beta\;$.

With 
$
\;\kappa\beta-2\xi X_{\rm nom}^{2} \ge \displaystyle c>0\;
$, $\mathbf{h}$ is \emph{contracting in the Euclidean metric} with rate at least $\displaystyle c$.
(Note that the rotation term $\omega_0 J$ drops out because $J^\top=-J$.)
By the Theorem~\ref{thm:sync}, this guarantees exponential synchronization.
\end{proof}

\begin{remark}
	Since contraction means exponential convergence, a contracting system exhibits a superior property of robustness by Theorem~\ref{thm:incstab_robust}. In a contracting system with rate $\lambda_{\max}(\mathbf{x},t)	\le -\displaystyle c < 0 $, any bounded deterministic disturbance  $\bar d$ yields only a bounded separation
	between trajectories: they remain exponentially attracted to each other and settle
	inside an invariant “error ball’’ whose radius grows linearly with the disturbance
	size (on the order of $\mathcal{O}(\bar d/ \displaystyle c $). Therefore, any error made in modeling or physical realization can be bounded.
\end{remark}

\paragraph*{Particular synchronized solution (existence and form)}

At the PCC, the complex voltage satisfies
\begin{equation}
	\label{eq:voKsh}
	\pha v_o \;=\; \beta\,\pha x_i\,\frac{\sum_{m} Y_m}{Y_\Sigma}
	\;=:\; K_{\rm sh}\,\beta\,\pha x_i,
	\qquad K_{\rm sh}\in(0,1),
\end{equation}
so when $\sum_{m}Y_m \gg Y_{\mathrm{net}}$ we have $K_{\rm sh}\!\approx\!1$ and
$\pha v_o(t) \approx \beta\,\pha x_i(t)$.

Substituting \eqref{eq:voKsh} into the node dynamics
\begin{equation}
	\label{eq:eachADF}
	\begin{aligned}
		\dot x_{\alpha,k} &= \chi\,x_{\alpha,k}-\omega_0 x_{\beta,k}
		-\kappa\!\left(\beta x_{\alpha,k}-v_{o,\alpha}(t)\right),\\
		\dot x_{\beta,k}  &= \chi\,x_{\beta,k}+\omega_0 x_{\alpha,k}
		-\kappa\!\left(\beta x_{\beta,k}-v_{o,\beta}(t)\right),
	\end{aligned}
\end{equation}
yields, in compact $\alpha\beta$–vector form with $J=\begin{bmatrix}0&-1\\1&0\end{bmatrix}$,
\begin{equation}
\begin{aligned}
&\dot {\pha x_{k}}
=\Big(\chi I+\omega_0 J-\kappa\beta(1-K_{\rm sh})I\Big)\pha x_k,\\
&\chi(\pha x_k)=\xi\!\left(2X_{\rm nom}^2-\|\pha x_k\|^2\right).
\end{aligned}
\end{equation}
Let $r_k:=\lvert \pha x_k\rvert$. Since $J$ is skew-symmetric,
the radial dynamics decouple:
\[
\dot r_k
=\Big(\xi(2X_{\rm nom}^2-r_k^2)-\kappa\beta(1-K_{\rm sh})\Big)\,r_k .
\]
Thus the synchronized periodic solution has constant amplitude $r^\star>0$ solving
\begin{equation*}
\begin{aligned}
&\xi\,(2X_{\rm nom}^2-{r^\star}^2)-\kappa\beta(1-K_{\rm sh})=0
\quad\Longrightarrow\\
&{r^\star}^2= 2X_{\rm nom}^2-\frac{\kappa\beta}{\xi}\,(1-K_{\rm sh}) .
\end{aligned}
\end{equation*}

Hence the particular synchronized trajectories are
\[
	\pha x_k^\star(t)=r^\star e^{j\omega_0 t},\qquad
	\pha v_o^\star(t)=K_{\rm sh}\,\beta\,r^\star e^{j\omega_0 t} 
\]
(provided the right-hand side of the amplitude expression is positive; otherwise
$r^\star=0$ corresponds to oscillator death). In the dominated case
$K_{\rm sh}\!\to\!1$, we recover
${r^\star}\to \sqrt{2}\,X_{\rm nom}$ and $\pha v_o^\star \approx \beta\,\pha x_k^\star$.

\begin{remark}
The particular solution $\mathbf{x}^\star(t)$ is \emph{globally attracting} by
contraction, so performance and limits (voltage magnitude, currents, sharing) can be
evaluated by the simple algebra of Proposition~\ref{prop:network} and the KCL relation \eqref{eq:finalsteady}—turning a
high-order nonlinear network into a tractable, decentralized calculation.
\end{remark}

\section{Validation}

\begin{figure}[!h]
	\begin{center}
		\includegraphics[width = 1\linewidth]{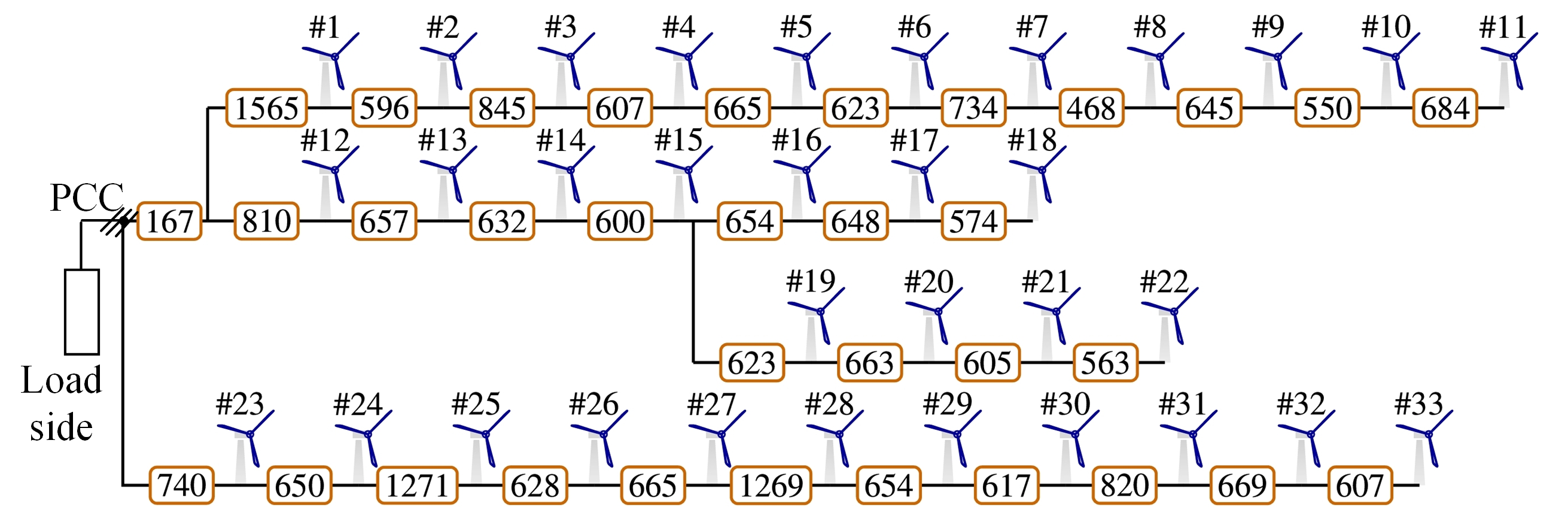}
		\caption{The layout of a real wind power plant containing 33 wind turbines and different transmission line impedances.}
		\label{fig:case-study}
	\end{center}
\end{figure}

\begin{figure}[!t]
	\begin{center}
		\includegraphics[width = 1\linewidth]{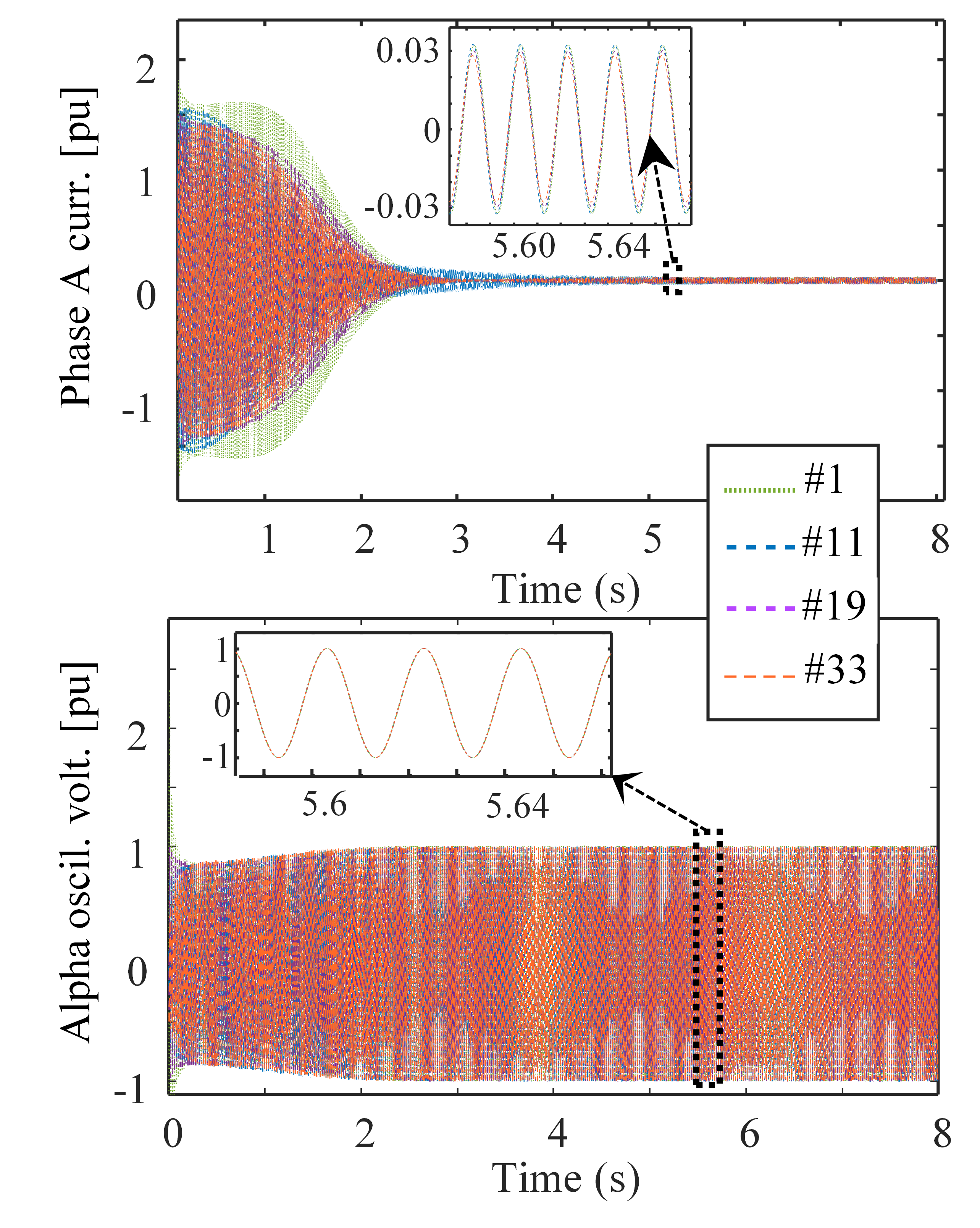}
		\caption{Case I: phase A current and alpha voltage $ x_{{\rm \alpha},k}$ of $\#$1, $\#$11, $\#$19 and $\#$33 inverters  during  start-up.  With random initial condition, the start-up transient shows that the multi-converter system is large-signal contracting: synchronized voltages and relatively identical current sharing.}
		\label{fig:case-study11}
	\end{center}
\end{figure}
\begin{figure}[!t]
	\begin{center}
		\includegraphics[width = 1\linewidth]{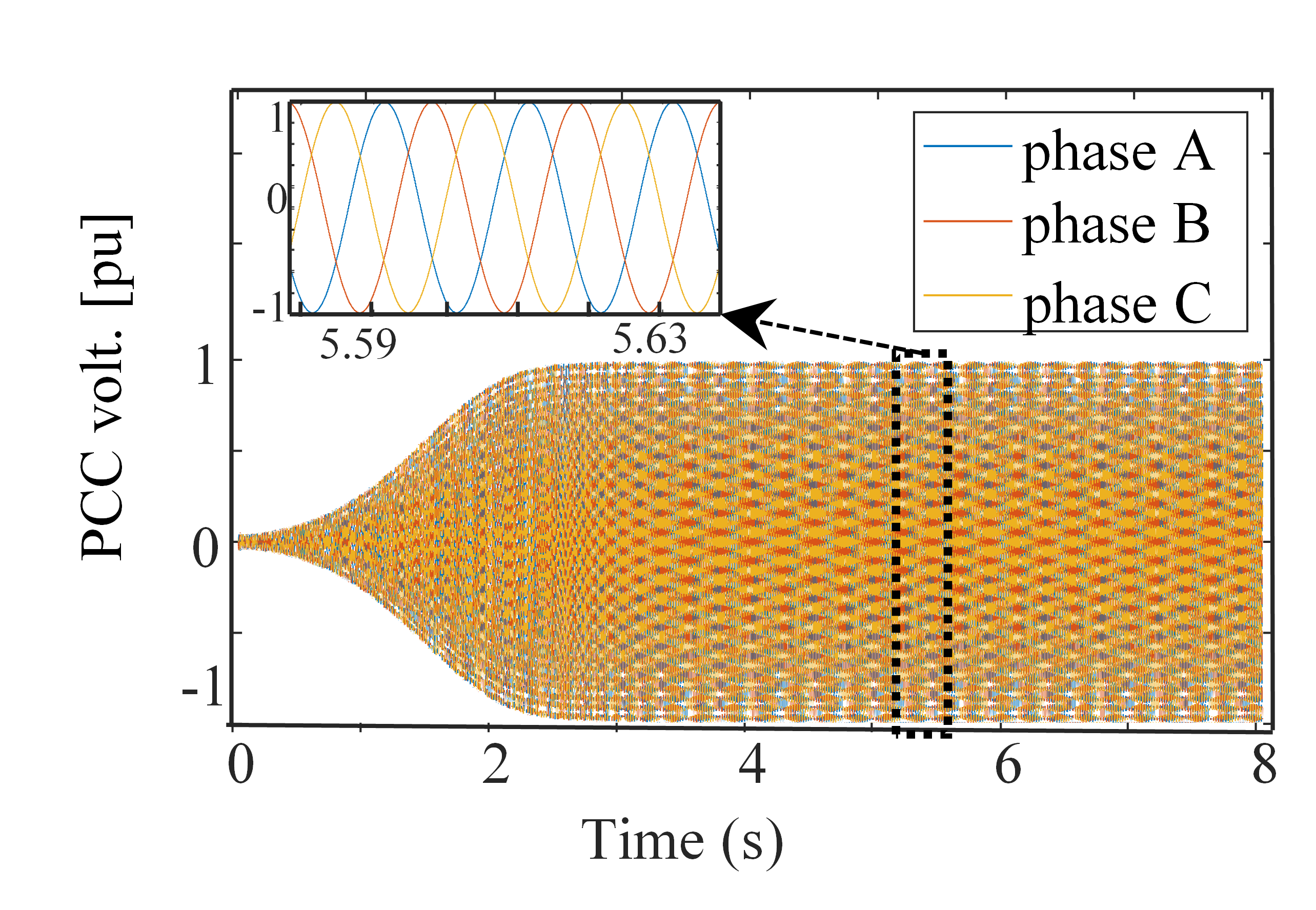}
		\caption{Case I: start-up transient of three-phase PCC voltage, $\pha v_o(t)$, under random initial condition. }
		\label{fig:case-study12}
	\end{center}
\end{figure}

\begin{figure}[!t]
	\begin{center}
		\includegraphics[width = 1\linewidth]{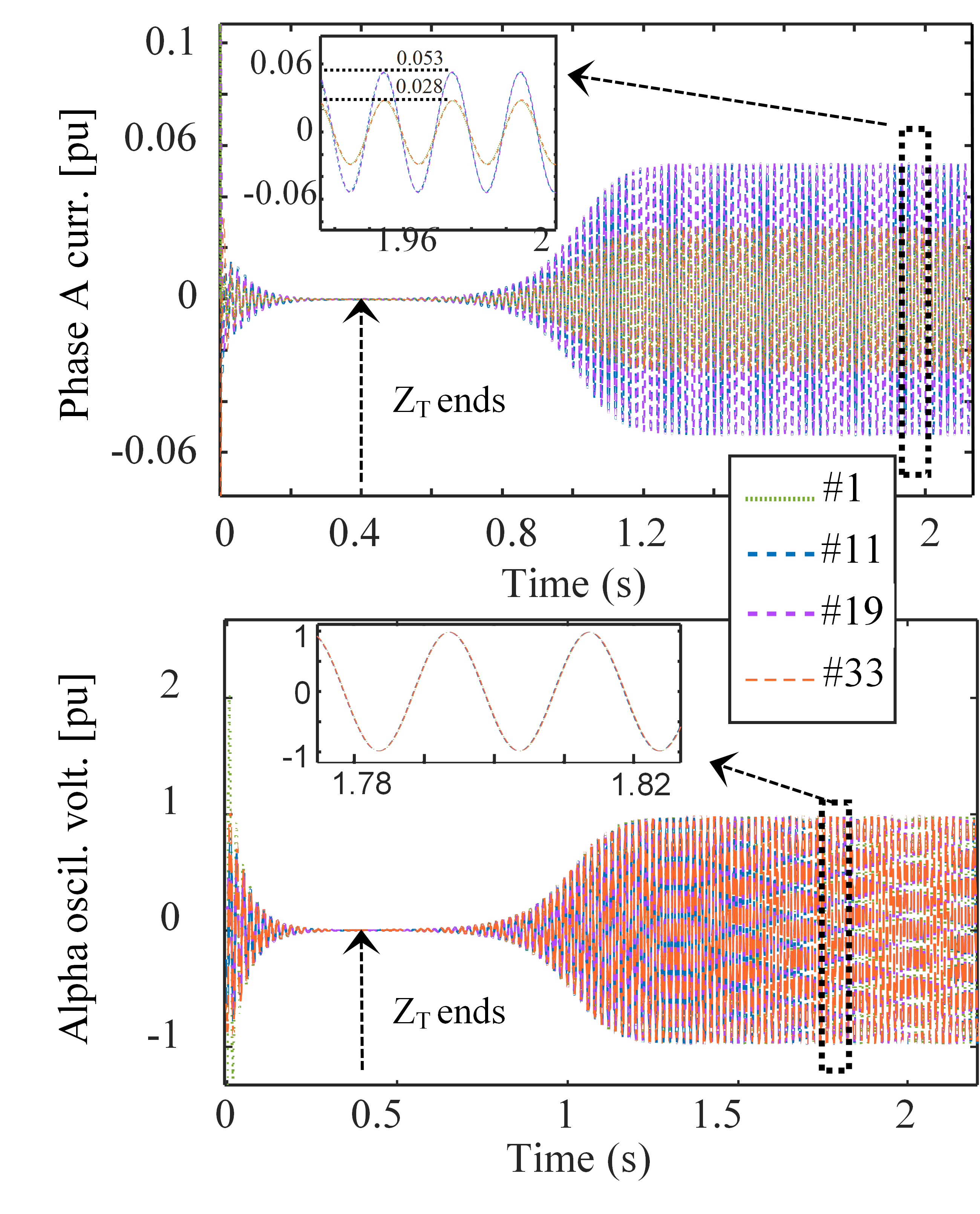}
		\caption{Case II: phase A current and alpha voltage $x_{{\rm \alpha},k}$ of $\#$1, $\#$11, $\#$19 and $\#$33 inverters during start-up. Applying a random chosen $Z_T$ from  $0$ second to $0.4$ seconds and accurate proportional current sharing,  $ \frac{I_{19}}{I_{33}}=\frac{I_{11}}{I_{1}}=\frac{0.053}{0.028} \approx \frac{20}{10.5}= \frac{Z_1}{Z_{11}}=\frac{Z_{33}}{Z_{19}}\, ,$ with large-signal voltage synchronization..}
		\label{fig:case-study21}
	\end{center}
\end{figure}

\begin{figure}[!t]
	\begin{center}
		\includegraphics[width = 1\linewidth]{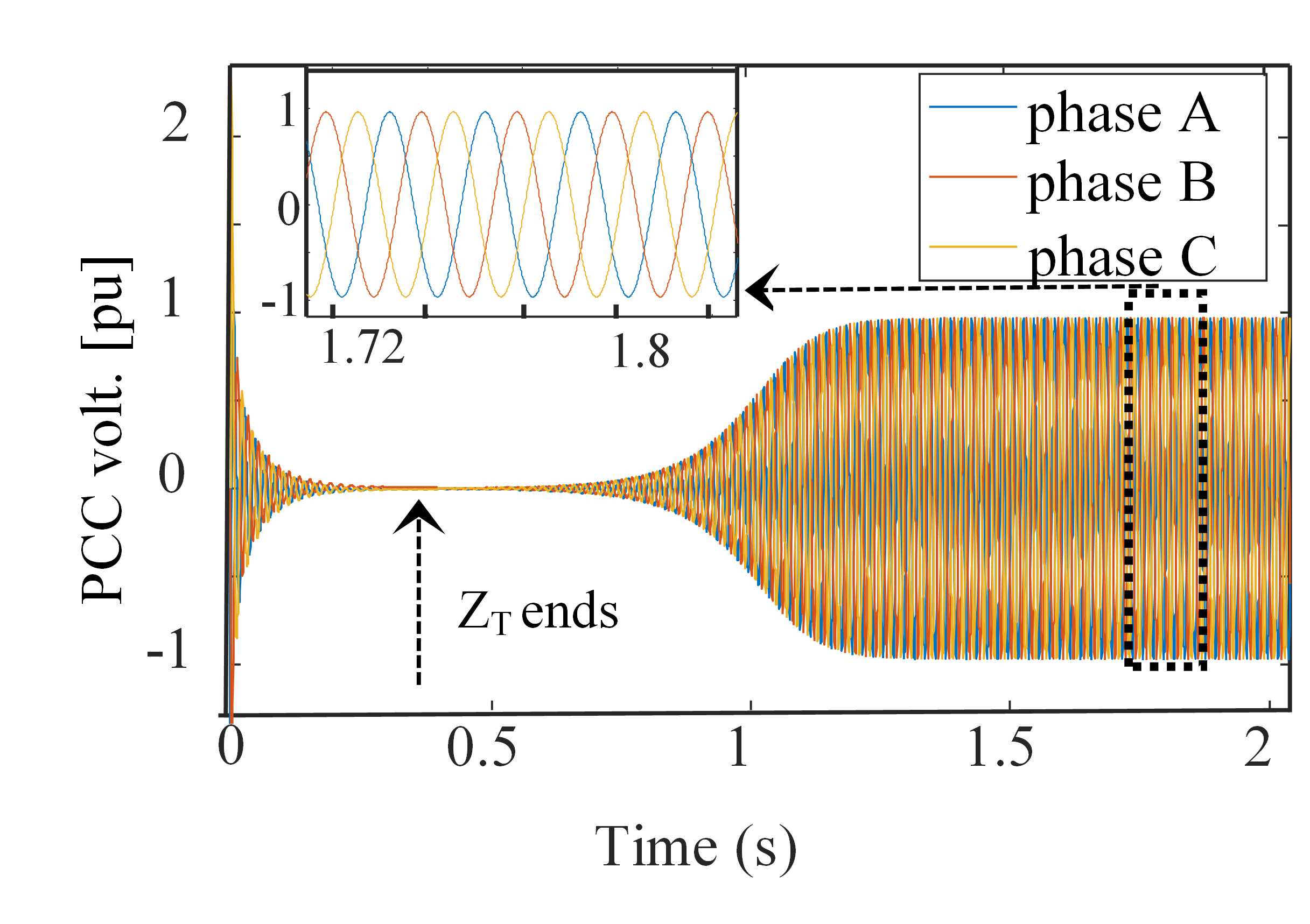}
		\caption{Case II:  start-up transient of three-phase PCC voltage, $\pha v_o(t)$, under random initial condition with a random chosen $Z_T$ from  $0$ second to $0.4$ seconds.}
		\label{fig:case-study22}
	\end{center}
\end{figure}

This part illustrates the contraction and virtual impedance in ensuring the transient stability and accurate current sharing of a wind power plant (WPP) which has 33 wind turbines. The previous work \cite{osti_2418742} offers only small-signal  stability without accurate current sharing function. The work \cite{Passivity2024} needs to know system-wide data and gives the stability condition under quasi-steady condition. The work \cite{Passivity2024} does not include  accurate current sharing function, either.   However, the proposed methods (contraction stability analysis in GFMI and contraction-based  controller with virtual-impedance) can certify \emph{global asymptotic stability with exponential convergence} and \emph{proportional current sharing} \emph{without} quasi-steady assumptions or system-wide data. The entire guarantee reduces to a simple, fully decentralized \emph{algebraic} inequality.
This case study is based on a real WPP layout \cite{li2017practical} depicted in Fig.~\ref{fig:case-study},  where all wind turbine converters are dVOCs. Load  is at $1.0$ pu with power factor $0$. Voltage setpoints are uniformly set to $1.0$ pu. Moreover, $\xi =10,\,2X_{nom}^2=1,\,\beta= 690*1.414/1.732\,\rm V$ are employed. The collector line parameter is $ Z_{pm}=0.1153\, + j\,\omega\,1.05\times10^{-3}\,{\rm \Omega}$/${\rm km}$.

To check the large scale synchronization, the local impedance without the virtual impedance for each inverter is designed as $0.75*Z_{pm}$ in Case I. The results of phase A current injected to load side and the alpha component of voltage in $\# 11 $, $\#19$, $\#1$ and $\#33$ dVOC are in Fig.~\ref{fig:case-study11}. The three-phase voltage at PCC node is in Fig.~\ref{fig:case-study12}. The random initial states are selected within $ \|\pha x_k(0)\|\le 1  $ in pu, except $\|\pha x_1(0)\|$ is selected as $10$ in pu to show  robustness of the n contracting inverters with $\;\kappa\beta-2\xi X_{\rm nom}^{2} \ge \displaystyle c> 543$. Although, at the beginning of huge states mismatching, at $5.6$ seconds, the n inverters have achieved synchronization, $ \forall\, i,j \in \{1,\dots,n\}:\;\lim_{t\to\infty}\|\pha x_i(t)-\pha x_j(t)\|=0$ and relatively identical current sharing. It is easy to show that the current sharing error appears because this setting does not satisfy the condition that load/local-impedance dominate the system’s voltage drop.

To ensure accurate current sharing, the local impedance with virtual impedance for each inverter is designed as $20*0.75*Z_{pm}$, except $\#11\, \text{and}\, \#19 $ are designed as $10.5*0.75*Z_{pm}$ in Case II. The results of phase A current injected to load side and the alpha component of voltage in dVOC in Fig.~\ref{fig:case-study21}. The three-phase voltage at PCC node is in Fig.~\ref{fig:case-study22}. It is predicted that this setting satisfies the domination condition.
Similarly, the random initial states are selected within $ \|\pha x_k(0)\|\le 1  $ in pu, except $\|\pha x_1(0)\|$ is selected as $10$ in pu to show  robustness of the n contracting inverters with $\;\kappa\beta-2\xi X_{\rm nom}^{2} \ge \displaystyle c> 543$. 
Additionally, to limit the current during start-up transient, the local impedance is randomly made 200 times of the local impedance said above until $0.4$ seconds. The additional local impedance is called $Z_T$ shown in Fig.~\ref{fig:case-study21} and Fig.~\ref{fig:case-study22}. It follows that the current sharing ratio of $\# 11 $ and $\#19$ to $\#1$ and $\#33$ is $ \frac{I_{19}}{I_{33}}=\frac{I_{11}}{I_{1}}=\frac{0.053}{0.028} \approx \frac{20}{10.5}= \frac{Z_1}{Z_{11}}=\frac{Z_{33}}{Z_{19}}\, ,$ both being about $1.90$ (relative difference $\approx 0.63\%$), which is predicted by \eqref{eq:finalsteady} caused by the shaping of virtual impedance.

The converters are average-valued, with fixed DC voltages and sufficiently fast inner-loop dynamics based on MATLAB/Simulink, where the complete control dynamics are included. The simulation results in Fig.~\ref{fig:case-study}(b) and (c) validate the transient stability of the contracting system featuring voltage synchronization and accurate current sharing simultaneously. Contraction stability analysis guarantees global exponentiation convergence from a black start with random initial condition in Fig.~\ref{fig:case-study}(a). With the help of contraction-based controller with virtual-impedance, the limited current start-up operation and accurate proportional current sharing is achieved in Fig.~\ref{fig:case-study}(b).


\section{Conclusion}

The contraction stability analysis in grid-forming inverters and contraction-based  controller with virtual-impedance are proposed in this paper. The transient stability of multi-converter systems is analytically studied with the grid-forming dVOC. The contracting dVOC controlled by contraction-based control with virtual-impedance in large-signal form makes itself noteworthy in stability guarantees. By leveraging both the contraction and virtual impedance of the dVOC node dynamics,  decentralized voltage synchronization and accurate current sharing conditions are developed to serve as a fast and effective tool for controller parameter tuning, stability guarantees, and large-scale integration of renewable energies. It can be extended to grid restoration, faults, unbalance, and grid connection analysis. It is hoped that a grid with virtual/real synchronous generators, droop-like and grid-following with current limiting inverters can all be analyzed in a contraction-theoretic framework for global synchronization and stability in future study.

\bibliographystyle{IEEEtran}
\bibliography{IEEEabrv,Bibliography}

\end{document}